\documentclass[sigconf]{acmart}
\usepackage{amsmath} % for align environment
\usepackage{bm} 
\usepackage{amsmath}
\usepackage{stackengine}

\usepackage{enumitem}
\usepackage{graphicx}
\usepackage{float}
\usepackage{balance} 
\usepackage{xcolor}
\usepackage{hyperref}
\usepackage{subcaption}
\AtBeginDocument{%
  \providecommand\BibTeX{{%
    \normalfont B\kern-0.5em{\scshape i\kern-0.25em b}\kern-0.8em\TeX}}}

\setcopyright{acmcopyright}
\copyrightyear{2021}
\acmYear{2021}
\acmDOI{10.1145/1122445.1122456}

\acmConference[ECML PKDD '21]{ECML PKDD '21: workshop on Bias and Fairness in AI}{September 13--18, 2021}{Bilbao, Spain}
\acmBooktitle{ECML PKDD '21: workshop on Bias and Fairness in AI}
\begin{document}
\settopmatter{printacmref=false} % Removes citation information below abstract
\renewcommand\footnotetextcopyrightpermission[1]{} % removes footnote with conference information in first column
\pagestyle{plain}

\title{Desiderata for Explainable AI in statistical production systems of the European Central Bank}


\author{Carlos Mougan}
\affiliation{%
  \institution{University of Southampton}
  \country{United Kingdom}}
\email{C.Mougan-Navarro@soton.ac.uk}

\author{Georgios Kanellos}
\affiliation{%
    \city{Directorate of General Statistics}
  \institution{European Central Bank}
  \country{}
}\email{George.Kanellos@ecb.europa.eu}

\author{Thomas Gottron}
\affiliation{%
    \city{Directorate of General Statistics}
 \institution{European Central Bank}
   \country{}
}\email{Thomas.Gottron@ecb.europa.eu}

\renewcommand{\shortauthors}{}

\begin{abstract}
Explainable AI constitutes a fundamental step towards establishing fairness and addressing bias in algorithmic decision-making. Despite the large body of work on the topic, the benefit of solutions is mostly evaluated from a conceptual or theoretical point of view and the usefulness for real-world use cases remains uncertain. In this work, we aim to state clear user-centric desiderata for explainable AI reflecting common explainability needs experienced in statistical production systems of the European Central Bank. We link the desiderata to archetypical user roles and give examples of techniques and methods which can be used to address the user's needs. To this end, we provide two concrete use cases from the domain of statistical data production in central banks: the detection of outliers in the Centralised Securities Database and the data-driven identification of data quality checks for the Supervisory Banking data system.\footnote{\textbf{Disclaimer:} This paper should not be reported as representing the views of the European Central Bank (ECB). The views expressed are those of the authors and do not necessarily reflect those of the ECB}

%In this work we state the desiderata of explainable AI in the statistical production system of the European Central Bank. We present a set of real explainable AI use cases where different explanations are needed. This work aims to make available to the public explainable AI needs.

\end{abstract}

%\begin{CCSXML}
%<ccs2012>
%   <concept>
%       <concept_id>10010147.10010257.10010258.10010259.10010263</concept_id>
%       <concept_desc>Computing methodologies~Supervised learning by classification</concept_desc>
%       <concept_significance>500</concept_significance>
%       </concept>
%   <concept>
%       <concept_id>10010147.10010257.10010258.10010259.10010266</concept_id>
%       <concept_desc>Computing methodologies~Cost-sensitive learning</concept_desc>
%       <concept_significance>300</concept_significance>
%       </concept>
%   <concept>
%       <concept_id>10010405.10010455.10010460</concept_id>
%       <concept_desc>Applied computing~Economics</concept_desc>
%       <concept_significance>100</concept_significance>
%       </concept>
%\end{CCSXML}

%\ccsdesc[500]{Computing methodologies~Supervised learning by classification}
%\ccsdesc[300]{Computing methodologies~Cost-sensitive learning}
%\ccsdesc[100]{Applied computing~Economics}


\maketitle

\section{Introduction}

%The establishment of machine learning in many areas from central banking, education, healthcare to public policy is raising questions about how to build fairness, explain the decisions, assure privacy and keep security on Artificial Intelligence systems.

Improvements in the predictive performance of machine learning algorithms have led to applications across many domains, including central banking, education, healthcare, and public policy. As machine learning becomes increasingly ingrained as a means to improve business efficiency and decision-making, it becomes crucial to establish processes for ensuring fairness, explainability, data privacy, and security. 

Explainability has become an important concept in legal and ethical guidelines for data and machine learning applications~\cite{intuitive_appeal}. The European Union General Data Protection Regulation (GDPR)\cite{gdpr} states that individuals have the right to receive an explanation for the output of an automated algorithm~\cite{gdpr_right}. Ensuring algorithmic accountability and transparency in machine learning models is a key ingredient towards understanding how decisions are made by the system. Understanding decisions is a prerequisite to detecting potential bias and ensuring fair and safe machine learning systems. 

The research community has made an enormous effort to design algorithmic methods for explainability. Various publications reconcile the information of all relevant papers, create explainable AI (xAI) taxonomies, define future opportunities and current gaps. \cite{guidotti_survey,paco_herrera_xai_taxonomi,exaplaining_explanations_ATI,explainabe_public_policy,explaining_explanations_overview_of_interpretability,xai_concepts,rudin2019stop}. It is worth noting that several authors have drawn attention to the \emph{need for industry and institutions to establish clear desiderata for their explanation needs}~\cite{xml_deployment,explainabe_public_policy}. Such desiderata are needed to assess if developed explainability methods address the requirements of real use cases.

In this paper, we leverage practical experience with machine learning applications in the domain of statistical production systems at central banks to present a user-centric classification of xAI requirements and needs. 
The user-centered approach aims to help both: (i) researchers to orient and describe their work along concrete requirements and (ii) practitioners to identify what kind of xAI challenge they face and which (technical) method might solve their problem. 
We illustrate the applicability of the desiderata classification scheme using two real-world use cases where xAI needs played an important role. For these two use cases, we map the requirements to the classification scheme and describe the technical algorithms chosen to solve the needs.

%We refer to real-world AI use cases from the domain of central banking to devise a systematic approach for algorithmic explainability. 

%Our approach is user-centric and aims to guide researchers and practitioners about how to face the xAI challenge at hand and what solutions are at their disposal. 

In this paper we make the following  contributions:
\begin{enumerate}%[label=(\roman*)]
    \item With our practical experience we identify a set of user-centric explainability needs that are generic and common throughout different use cases.

    \item We provide concrete examples of use cases that make the classes of desiderata more practical, aiming to help identify potential research gaps.

\end{enumerate}

The rest of the paper is organized as follows: in Section~\ref{sec:related} we introduce related work and outline differences with our contribution. Section~\ref{sec:taxonomy} introduces the user-centric classification with definitions of the explainability needs together with the user roles involved in the process. In section~\ref{sec:useCases} we describe concrete use cases with their proper introduction, state desiderata, and map business requirements to the classification scheme and algorithmic solutions. Finally, in Section~\ref{sec:conclusions} we summarise the main conclusions of the paper and outline possible future work.
\section{Related Work}\label{sec:related}
Research on xAI has gained a lot of attention \cite{peeking_bb} and a growing number of research papers on the topic is being published in conferences and journals \cite{xai_concepts,guidotti_survey}. According to the literature, papers in xAI methods might be suffering from two issues:
\begin{enumerate}[label=(\roman*)]
    \item A lack of a rigorous definition and evaluation methods for explainability~\cite{rigorousXAI,mythos_interpretability}
    
    \item Explainability methods are introduced as general-purpose solutions and do not directly address real use cases or a specific user audience~\cite{explainabe_public_policy,bewareInmates}
\end{enumerate}

The challenge of evaluation methods for xAI is typically addressed with user studies. 
Few recent papers evaluate explanation methods in a very specific real-world task with real users~\cite{how_can_I_choose_an_explainer},  but they do not offer a generalisation to other use cases. More often, previous work found in the literature relies either on theoretical problems or toy datasets \cite{ATI_taxonomy}.
Accordingly, the effectiveness and usefulness of theoretical research papers on real-world applications are still unclear~\cite{bewareInmates,mythos_interpretability}.
There are few real explainable machine learning use cases where a full xAI methodology is presented addressing intended users and tasks. 

The challenge of mapping explainability methods to concrete use cases is addressed by  Bhatt et~al.~\cite{xml_deployment}. They critically examine how explanation techniques are used in practice by interviewing various organizations on how they employ explainability in their machine learning workflows. They state that there is a need for the industry to define clear desiderata for explainability, with clear goals and intended users. Our work aims to address that gap by defining needs, users and showcasing them with two use cases.
Amarasinghe et~al.~\cite{explainabe_public_policy}, present a position paper where they aim to define xAI roles and intended users for the public policy domain. Our work extends their research by redefining the taxonomy with a more user-centric approach and making use of two particular use cases to identify more specific explainability needs.

To the best of our knowledge there is no previous work done on the desiderata for explainable machine learning in use cases for statistical production systems, nor on stating user-centric xAI requirements for a public policy institution.

\section{User-centric Classification of Desiderata for Explainability Needs}\label{sec:taxonomy}

While working on several central banking projects involving machine learning we observed recurrent patterns of where needs for explainability arose. This experience motivated us to structure and classify the needs and requirements for explainability. The needs are described in a user-centric way in order to identify stakeholders and their specific requirements. This user-centric approach also allows system designers to work with \emph{personas}\footnote{We refer to the concept of personas as it is used in approaches like design thinking.} to reflect specific needs for xAI.

In this section we introduce and describe our classification of generic explainability needs. Furthermore, the entries in the classification link xAI needs with possible methods to address them. We do not intend to provide an in-depth survey on existing work\footnote{Please refer to some of the paper referenced in related work for comprehensive surveys}. Rather we intend to give examples of existing work that meet the xAI requirements for some use cases we encountered.

The needs are synthesized from practical use cases of machine learning at the European Central Bank (ECB) in the context of statistical data production. We will provide two concrete use cases to fill the classification with life in Section~\ref{sec:useCases}. While the use cases will serve as examples on how to use the classification, the desiderata reflect generic user needs which in our view cover most relevant use cases of xAI.

\subsection{Users}

As stated above, we came across several generic user roles which help to classify the needs for solutions of explainable and responsible AI. A key question driving this classification is \emph{Who needs an explanation of an AI method?} This helps to clearly define and distinguish different desiderata for explanations. 

The following profiles define users that can potentially interact with a machine learning system all of whom have quite different needs for explainability:

\begin{enumerate}[label=(\roman*)]
    \item \textbf{Data scientists and AI engineers}: This role corresponds to members of the team who build, model and run a machine learning application. This type of user has technical expertise but does not necessarily have business expertise. They are in charge of the full life cycle of the machine learning application, from development to maintenance in production.

    \item \textbf{Business experts}: Users with this role provide the use case and domain expertise for a machine learning solution. They define the business activity or process which is supported by the AI solution. In our case, they are finance, economics and statistics experts from the European System of Central Banks who act or intervene in business processes based on the recommendations of the models. This type of user might not have a technical background and is not a machine learning expert.
    
    \item \textbf{High stake decision makers}: This type of user determines whether to use and incorporate a machine learning model in the decision-making process. They typically have a management position, a high-level understanding of the business objectives and a responsibility to deliver value. They need to understand and assess the potential risk and impact of incorporating the machine learning model into production.
    
    \item \textbf{End users}: Users which are affected by or make use of the final results belong to the group of end users. The knowledge and potential expertise of this user group varies significantly. There might be cases where the group of end users actually overlaps or is even identical to the group of \emph{business experts}, e.g. when the machine learning solution is primarily serving internal business processes. Examples of end users in our domain are the business areas or even the general public making use of data compiled by the Directorate General Statistics at the ECB.
\end{enumerate}

\subsection{Building trust}

Trust is essential for the adoption of a machine learning application. Depending on the user, the meaning of trust and the way to obtain it differ.
In this section we analyse the different types of trust which need to be established for the four user types.

%\subsubsection{Trust by the Data Scientist}

\emph{Data scientists and AI engineers} are responsible for the whole life cycle development of the machine learning model~\cite{mle}. The engineering phase of the cycle includes tasks like data exploration, deciding on an evaluation scheme, feature engineering, selecting and training a model as well as evaluating the results in a systematic way to ensure performance in a production environment.

During this phase, the \emph{data scientist} has to ensure that meaningful features are selected, that there are no data leakages and that spurious correlations are addressed. This can be achieved by performing sanity checks and understanding the model output through the use of feature relevance explainers such as SHAP~\cite{shapley}, tree Explainer~\cite{shapTree}, LIME~\cite{ribeiro2016why,ribeiro2016modelagnostic}, and local statistical aggregations~\cite{statisticallearning}. 

%The needs for explainability slightly shifts once a model is entering its production phase. As part of monitoring the model performance it is important to understand if the model still behaves as designed in the engineering phase. This means to understand if the quality of the model remains the same, if the same features have an influence on the models conclusions and which factors might lead to a deterioration of the model's performance.

%\subsubsection{Trust by the Business Expert (Data Quality Manager)}

\emph{Business experts} have knowledge about business processes and often have practical experience in how the task for the machine learning systems should be performed. In certain cases the \emph{business experts} might even have been responsible to actually perform the task that is to be taken over by a machine learning solution. Hence, from their perspective the aspects that need to be addressed to generate trust in a machine learning algorithm are the following: (\emph{i}) adapting to a changed environment (\emph{ii}) understanding the decisions of the algorithm.

The disruption caused by the introduction of machine learning in the process needs be mitigated in order for the experts to have enough time to understand, test and embrace the new method. To this purpose, the predictions or recommendations of the machine learning algorithm can be introduced and transparently explained in parallel to existing measures and processes so that the experts can familiarise themselves with the new functionality. Once it becomes clear to them that by making use of machine learning they can achieve equal or higher levels of effectiveness and efficiency, transition is simpler. 

The second issue that needs to be addressed is the explainability of the results. As part of their business processes, \emph{business experts} might be required to be able to understand the reasons or facts that led to the recommendations and decisions produced by the algorithm. The decision-making process might need to be assessed internally or even externally e.g. in the context of audits.

It is worth noting that the technical level of this type of user is generally lower than that of the data scientist, so normally a less technical explanation or even an \emph{explanation of an explanation} is required. This can be achieved via feature attribution techniques such as SHAP and LIME. It helps immensely in the trust building process if the criteria for the decisions reflect at least partially the prior knowledge of the \emph{business experts}.

%\subsubsection{Trust by senior management}
For \emph{high stake decision makers} the need to understand machine learning algorithms reaches yet another, even less technical level. They take high-stake decisions and thereby responsibility for the adoption of AI solutions in a production environment. They do not need to understand the machine learning solution on the basis of individual decisions, but on an aggregated level. They need to understand the opportunities in terms of gains in efficiency and effectiveness, ideally using quantifiable measures. At the same time they need to assess risks that come with the adoption of a solution. The risks relate to the probability of errors and the impact of such errors on various levels (operational, competition, reputation, etc.). In particular, it might be required to ensure that an AI system is not biased and non discriminatory. Hence, explainability needs to be mapped to different concepts, which allow for assessing a solution\footnote{We hypothesise that this is probably one of the most challenging fields for xAI research.}.

Furthermore, \emph{high stake decision makers} might want to make sure that solutions for explainability for individual AI based decisions are available. This is managerial and operational prudence to ensure that the needs and requirements of the \emph{business experts} are satisfied (see above).

%\subsubsection{Trust by the data users}

\emph{End users} are normally not involved in the data processing part of the process, which means that the algorithms used are not known to them. Nevertheless, ensuring that they continue to trust the data after the introduction of machine learning is a crucial part of the data workflow.
This comes down to ensuring they trust the outcome of the algorithm either when it affects the end users themselves or when it is used as a service.
%This might be related to the end users to trust the outcome itself as a service they consume or because the outcome of a machine learning directly affects the end users themselves.

\emph{End user} trust is obtained through consistently achieving high output quality as well as clear and transparent communication. For example, in the context of statistical publications which have certain timeliness and continuity in the data, sudden variations might distress users. Therefore, the introduction of a new ML algorithm in production must be validated thoroughly and in case it introduces significant changes in the data output of the system these must be clearly communicated in advance to the users so that they understand the underlying reasoning.

\subsection{Gaining Knowledge through Machine Learning}

Besides very high prediction performance, machine learning can enhance data understanding. The aforementioned high performance is achieved through the "discovery" of patterns or connections that have not be utilised by other approaches. Understanding these patterns and what they mean can offer important new insights~\cite{peeking_bb}. Such insights are of particular relevance for \emph{business experts}. They may use the discovered patterns together with their business experience on new use cases, or to obtain deeper domain knowledge. 

The insights, however, need to be presented in an interpretable way. The patterns have to be communicated to the \emph{business experts} in a way that they can make sense of it. The requirements for this type of explainability are different than for the understanding of a model for the originally intended task. A wide range of different explanations, representations and visualisations might be needed to stimulate a creative cognitive process with the \emph{business experts}.

The evaluation of the pedagogical aspect of xAI techniques for machine learning models remains as a future avenue of research.
In Section~\ref{sec:useCases} we will provide an example from the domain of central banking, where gaining knowledge was one of the primary objectives for a machine learning project.

\subsection{Model monitoring}
Model monitoring aims to ensure that a machine learning application in a production environment displays consistent behavior over time. Monitoring is mainly performed by the \emph{data scientist and AI engineer} and is crucial, as a drop in model performance will affect all the users. Two common challenges in model monitoring are $(i)$ distribution shifts in the input data that can degrade model performance $(ii)$ changes in the machine learning algorithm due to a model retraining that can alter the individual explanations for decisions.
%As a consequence, flawed and under-performing machine learning pipelines can go undetected even in core software companies~\cite{rules_of_mle,continual_learning}.

The task of \emph{monitoring distribution shifts} can be formalised as follows:
Given source dataset $D_S$ and test dataset $D_T$, consisting of input $x$ and targets $y$, drawn from underlying distributions $p_S(x,y)$ and $p_T(x,y)$ respectively, we aim to detect changes in the distribution in $p_T$~\cite{continual_learning}. This phenomenon is known as distribution shift or concept drift~\cite{datasetShift,continual_learning} and is particularly difficult to notice\cite{monitoring}. 

%To address this issue we consider two different types of model monitoring where aspects of xAI are of relevance:

%Traditional statistical methods that are both simple and scalable can be used such as: Population Stability Index (PSI),  Kolmogorov-Smirnov or Kullback-Leibler divergence to measure statistical distance between $p_S$ and $p_T$. These methods are often limited to real valued data, low-dimensions and require certain probabilistic assumptions~\cite{continual_learning}. An approach suggested by Lundberg et~al.~\cite{shapTree} is to monitor the SHAP value contribution of input features  over time together with decomposing the loss function across input features in order to identify possible bugs in the pipeline.

The challenge from an xAI perspective is how to assess such changes from the perspective of the \emph{business expert} and the \emph{AI engineer}. \emph{Business experts} need to be able to identify if the distribution shift can be explained from a business perspective. Hence, they need to be able to identify possible causal relationships between the change in the data and exogenous events. On the other hand, the \emph{AI engineer} needs to assess the impact of the change on the model's performance and potentially intervene.

\emph{Monitoring changes in the explanation} addresses another phenomenon. 
The performance of machine learning models degrades over time. In order to maintain high performance, models are retrained using previous and new input data. This continual learning process can lead to changes in individual explanations for decisions through time. As an example, the explanation offered by the classifier $f_C(x)$, for the decision taken for the $i$-\textit{th} instance at a time $t_0$, might differ compared to the explanation offered at time $t_n$ for the same instance after model retraining.

\begin{equation}
    E_{t=0}(f_C(x_i)) \neq E_{t=n}(f_C(x_i))
\end{equation}

This change due to the continual learning process, increases the risk of contradictory explanations for the same instance. Local explanations are prone to this kind of risk, due to the fact that small changes in the neighborhood of the instance can lead to a different reasoning from the model once retrained.

Both, \emph{business experts} and \emph{AI engineers} need to be able to detect and understand this phenomenon. For the \emph{business experts} such a change might indicate a change in the business logic, for the \emph{AI engineer} it might indicate to review the modelling process and design decisions taken during the development phase. Depending on the use case, even \emph{high stake decisions makers} may need to be aware of such changes, as inconsistent explanations for the same setting may pose a risk of unfair algorithmic decisions.

%An effective way of addressing distribution shifts is to monitor the contribution of each individual feature to the output loss with feature relevance explainers such as Shapley~\cite{shapley}, tree Explainer~\cite{shapTree} or Lime~\cite{ribeiro2016why,ribeiro2016modelagnostic}. In this case the purpose of the xAI solution is to actually identify the change in model behaviour as indicator for enabling model monitoring.

%Once the model is deployed monitoring the loss function is unfeasible. An approach suggested by Lundberg et~al.~\cite{shapTree} is to monitor the SHAP value contribution of input features  over time together with decomposing the loss function across input features in order to identify possible bugs in the pipeline.

%Lundberg et al. \cite{shapTree} provides examples where local model monitoring together with the aforementioned decomposition by shap values can be used to identify potential bugs in the pipeline.

\subsection{Actionable Insights}

Understanding a machine learning algorithm is usually not an end in itself. The explanation offered through this understanding supports business processes and leads to actionable insights. Such insights enable the \emph{business expert} to understand how to change a decision by manually intervening in the data. 
For instance, when data is identified to belong to a certain class, providing a set of actionable changes that would lead to a different decision can assist an expert in correcting or modifying the data. 

Counterfactual generation aims to address this issue by proposing to the \emph{business experts} \emph{the minimal feasible change in the data in order to change the output of the algorithm}. Such a process enhances the understanding of the experts (and might further foster trust in the system). We formulate the problem following the counterfactual recourse formulation by Utsun et~al.~\cite{utsun_linear_recourse,ActionableRecouserLinear,AlgRecourse_Counterfactual_Interventions} and the open source python package DICE \cite{dice} that quantifies the relative difficulty in changing a feature via feature weights.

%\begin{equation} \label{eq:counterfactual}
%\delta^* \in \stackunder{argmin}{$\delta$} \quad cost(\delta: x^{F}) \quad s.t. \quad h(x^{CFE}) \neq h(x^F)
%\end{equation}
%\begin{subequations}
%\begin{align}
%x^{CFE} = x^F + \delta\\
%x^{CFE} \in P, \delta \in F
%\end{align} 
%\end{subequations}

%Where cost $(\cdot; x^F): X x X \rightarrow R_+$ is a user-specified cost that encodes preferences between feasible actions from $x^F$, and $F$ and $P$ are optional sets of feasibility and plausibility constraints, restricting the actions and the resulting counterfactual explanation respectively~\cite{AlgRecourse_Counterfactual_Interventions}. The feasibility constraints, as introduced in~\cite{ActionableRecouserLinear,AlgRecourse_Counterfactual_Interventions} aim at restricting the set of features that the individual may act upon.

\subsection{Fostering Explanations through Simple Models}\label{sec:simple_models}

Copying \cite{copies,copying_irene_sampling} or distilling \cite{distillation}  machine learning models can greatly contribute to model explainability. Overly complex models tend to be difficult to explain~\cite{mle} and can become unaccountable~\cite{rudin2019stop}. Model agnostic copies with a simple model might be able to achieve global explainability~\cite{unceta2018global} which can be useful to build trust and gain knowledge by the \emph{business expert}. Furthermore, in some deployment scenarios involving incompatible research and deployment versions~\cite{copies}, copying the ML model can ease the deployment task for the \emph{data scientist}.

We now define what is copying a machine learning classifier \cite{copies,copying_irene_sampling,irene_enviromental}. We use the term original dataset to refer to a set of pairs $X = {(x_i, t_i)},i = 1, ..., M,$ where $x_i \in R_d$ is a set of $d$-dimensional data points in the original feature space D and $t_i \in {1, . . . , K}$ their corresponding labels. We define the original model, $f_O: X \xrightarrow{} t$, as any model trained using $X$. Our aim when copying is to reproduce the behavior of the original model employing a copy, $f_C$, such that $f_C(x) = f_O(x),  \forall x \in  D$. To do so we use a two-step process. Firstly, we rely upon the generation of a synthetic target, $y_j = f_O(X)$, where $y_j$ is the target label associated with each individual sample from the original data distribution $X$. Secondly, we use the synthetic target points to train a copy, $f_C$, whose decision function reproduces $f_O$ to the extent that it can be used to substitute it.

%deployment of a model is the complexity of the selected machine learning model. Overly complex models tend to be difficult to explain~\cite{mle} and can become unaccountable~\cite{rudin2019stop}. Furthermore, they more pose challenges for the infrastructure to run the model. Model agnostic copies and knowledge distillation might be solution for this challenge~ \cite{distillingTree,distillation}.

%\subsection{Improving decision-making - HCI approach}
%\subsection{Human - Computer Interaction}
%Using explanations to help users to take better decision in difficult cases. This should achieve better performance than individually human and the algorithm.

\section{Use Cases}\label{sec:useCases}
Providing statistical data and information of the highest quality is a core task of the European System of Central Banking (ESCB). Assuring the quality of data in statistical production systems is crucial as it is used in the decision-making process. Statistical production systems rely on domain experts with an outstanding understanding of the data. These experts ensure that the information used in the compilation of statistical products is of the highest quality based on their expertise.

The benefit of using expert knowledge for quality assurance of statistical products is two-fold: $(i)$ all the knowledge and expertise of the domain experts are reflected in the final data and $(ii)$ the overall confidence about the final product increases on the side of the producers and consumers of official statistical data. However, as the volume and granularity of the data increase, this process becomes progressively difficult to maintain. It is almost impossible to manually assess the quality of granular data without substantially increasing the number of experts.

Over the years various approaches have been developed to support and automate quality assurance procedures. However, these approaches have certain limitations:
\begin{itemize}
    \item \textbf{Data aggregation:} Assessing the quality of the data on an aggregated level, condenses the information that needs to be checked to a manageable amount. The disadvantage of this approach is that aggregation might hide granular data quality issues on the level of individual observations. The risk of obfuscation depends on the type and level of aggregation used.
    \item \textbf{Static quality rules:} Another way to ensure data quality for a big data scenario, is using a fixed set of rules to identify potential quality issues. These rules are based on expert knowledge and formalize certain aspects of their domain knowledge. Such rules flag individual records that, for instance, exceed a predefined threshold. The flagged observations typically represent only a small fraction of the entire data set, allowing the field experts to focus their analysis on a limited amount of data. The shortcoming of this approach is that relying on a fixed set of static rules is not flexible enough to capture dynamics in data nor does it automatically adjust to new insights provided by users. 

\end{itemize}

%Both of the aforementioned approaches can be useful, but they have clear limitations. Aggregation can hide granular data quality issues which could become visible if a different type of aggregation is selected by the data users while relying on a fixed set of rules is not flexible enough to capture changes in the data nor does it adjust to new user input. 

To maintain high standards of quality assurance, new approaches that address these limitations are investigated. This section illustrates two applications of machine learning solutions in the area of central banking statistics and how different user groups and needs for xAI have been addressed.

\subsection{Centralised Securities Database}
The Centralised Securities Database (CSDB)~\cite{csdb_asier} is a securities database with the aim of holding complete, accurate, consistent, and up-to-date information on all individual securities relevant for the statistical and, increasingly, non-statistical purposes of the ESCB.
Ensuring quality in such a system is challenging given the amount of data that needs to be monitored. So far, it has been achieved through a combination of data aggregation and static quality rules. In addition, \emph{business experts} with the role of Data Quality Managers have their internal measures for assessing data quality and intervening in case of issues. 

\subsubsection{Machine learning application}
The objectives of our approach were \cite{csdb_ml}:
\begin{enumerate}[label=(\roman*)]
\item Identify records in the dataset where there is a high chance that manual intervention is needed.
\item Communicate to the \emph{business experts}r which is the field that is most likely to be incorrect.
\item Rank these suggestions to minimize the number of false-positives in the top recommendations.
\end{enumerate}

\subsubsection{Building Trust}
Despite the careful testing and calibration of the machine learning process by the \emph{data scientists}, \emph{end users} of the data can identify potential data issues that have not been raised by the algorithm. These issues are communicated via the Data Quality Managers (DQM) to the statistical production team (comprising of \emph{business experts} and \emph{data scientists}) responsible for the CSDB data quality \cite{csdb_ml}.

\begin{figure}[tb]
\begin{subfigure}{.5\textwidth}
  \centering
  % include first image
  \includegraphics[width=.8\linewidth]{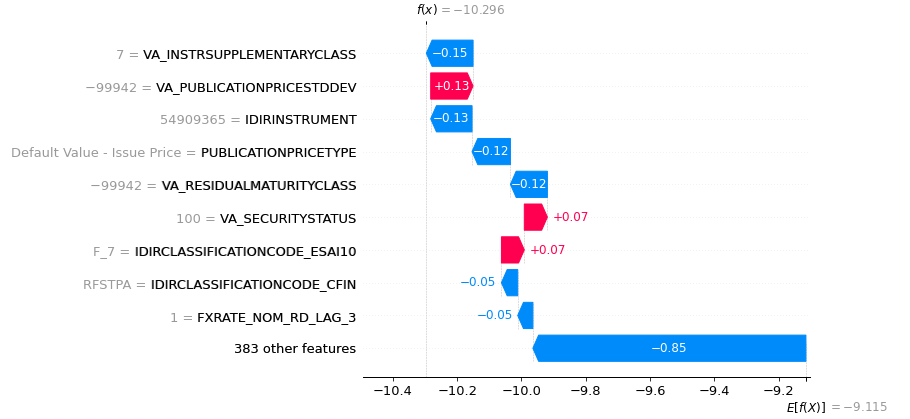}
  \caption{Local feature importance for a given instance\cite{shapley}}
  \label{fig:sub-first}
\end{subfigure}
\begin{subfigure}{.5\textwidth}
  \centering
  % include second image
  \includegraphics[width=.8\linewidth]{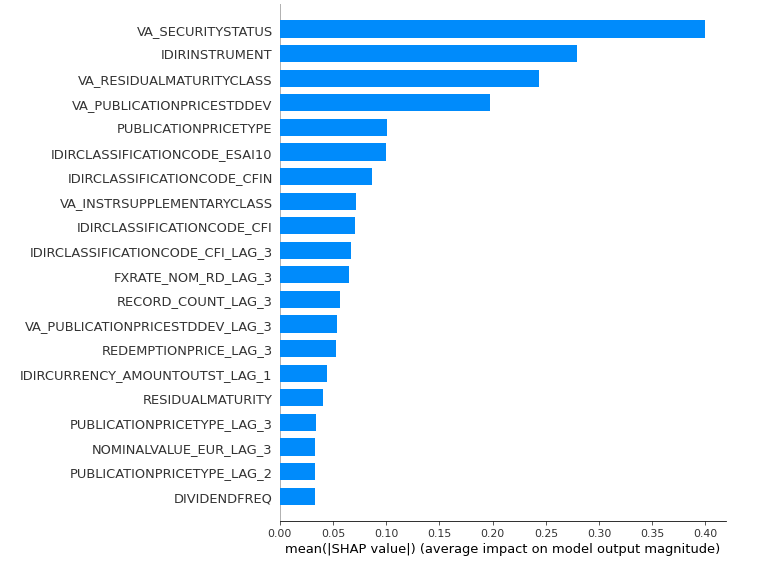}
  \caption{Global Feature Relevance}
  \label{fig:sub-second}
\end{subfigure}
\caption{Shapley feature values by TreeExplainer \cite{shapTree}}
\label{fig:fig}
\end{figure}

The first step for the statistical production team is to verify that indeed the issues identified are not flagged by the algorithm. The next step is to investigate the reasoning behind this choice from the perspective of the algorithm by determining: $(i)$ similar instances in the training dataset \cite{case_based_explanation}, $(ii)$ which features contributed to the decision \cite{shapTree,ribeiro2016why} (cf. Figure~\ref{fig:fig}) and $(iii)$ local decision rules to better understand the model logic \cite{lore,Ribeiro2018AnchorsHM}.

%\begin{figure}
%    \centering
%    \includegraphics[width=.8\linewidth]{images/lime.png}
%    \caption{Local approximation for feature importance by Lime \cite{ribeiro2016why}}
%    \label{fig:my_label}
%\end{figure}
\subsubsection{Actionable Insights}

\begin{figure}[bt]
    \centering
    \includegraphics[width=.6\linewidth]{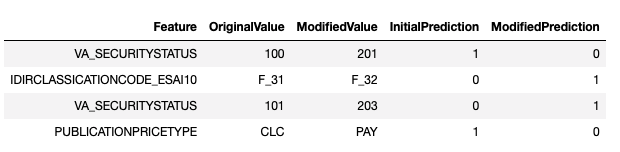}
    \caption{Set of counterfactual decisions generated \cite{lore,dice}}
    \label{fig:counterfactual}
\end{figure}

Following the formulation from Utsun et~al.~\cite{utsun_linear_recourse} and  the available open source implementation \cite{dice}, we quantified the relative difficulty in changing a feature through the application of weights to the  counterfactual explanations algorithm. For instance, in our case,  recommendations should not ask the \emph{business expert} (DQM) to modify the country in which a financial instrument was issued or change the issue date to a time before the creation of the issuing company (cf. Figure~\ref{fig:counterfactual}).

%\subsubsection{Gain Knowledge}
%During our work, after several models and feature attribution visualizations, we noted that what was happening with our data was not what we expected but users were doing other things.

\subsubsection{Model Monitoring}
Detecting when the underlying distribution of the data changes is paramount for this use case, since failing to predict outliers or errors in the data will lead to a drop in the trust of the machine learning model. Also, the risk of having an incoherent explanation through time caused by the continual learning process is utterly important, a discrepancy will lead to a decrease of trust by the \emph{business experts}. 

An approach suggested by Lundberg et~al.~\cite{shapTree} is to monitor the SHAP value contribution of input features over time together with decomposing the loss function across input features in order to identify possible bugs in the pipeline.

\begin{figure}[tb]
    \centering
    \includegraphics[width=.6\linewidth]{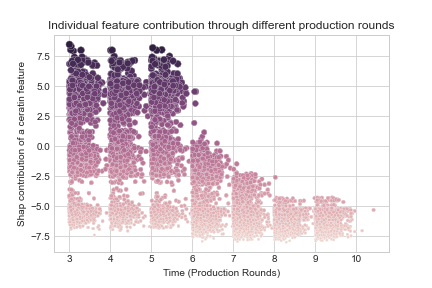}
    \caption{Simulated covariate distribution shift detected by local attribution of Shapley values in one of the descriptive features }
    \label{fig:local_monitoring}
\end{figure}

In Figure~\ref{fig:local_monitoring}, we can see the local contribution of SHAP values through time for a continuous feature. The drop of Shapley contributions denotes a possible distribution change for a given feature or a change in the model's inner working that needs human assistance for further evaluation of performance and explainability ~\cite{shapTree}. 

\subsubsection{Fostering Explanations through Simple Models}
Following the notation of paragraph \ref{sec:simple_models}, for our case the original model $f_O(X)$ is a catboost classifier~\cite{dorogush2018catboost,prokhorenkova2019catboost} which is a gradient boosting model~\cite{gbm} that often achieves state of the art results in many different types of problems \cite{supervised_learning_algorithms,gbm_travel} and the copied models $f_C(X)$ are a scikit-learn~\cite{scikit-learn} decision tree classifier and a Generalized Linear Model (GLM). The simpler models are of slightly inferior quality  (cf. Table~\ref{tab:table-name}). However, having two simple models helps to improve the overall global explainability for the \emph{data scientist} and to simplify deployment.

\begin{table}[bt]
\centering
\caption{Catboost is the original model $f_O(X)$ and the Decision Tree and the Generalized Linear Model the  copied classifier $f_C(X)$}
\label{tab:table-name} 
\begin{tabular}{lccc}
\toprule
          & Catboost & Decision Tree & GLM \\ %\hline
\midrule
AUC       & 0.816    & 0.771         & 0.741        \\
Precision & 0.805    & 0.741         & 0.685        \\
Recall    & 0.833    & 0.824         & 0.733   
\\
\bottomrule
\end{tabular}
\end{table}

\subsection{Supervisory Banking data system}
The Supervisory Banking data system (SUBA) is the ECB's system to collect supervisory data on credit institutions. 
The data collected in SUBA is in line with the reporting requirements defined by the European Banking Authority (EBA).
SUBA implements all validation rules defined by the EBA to assess data quality. 
These validation rules implement logical and business-motivated checks of consistency and correctness for the reported data.

However, supervisors and \emph{business experts} at the ECB may define additional \emph{plausibility checks} to further ensure the quality of the data.
So far, the additional \emph{plausibility checks} were defined manually based on domain expertise.
In an internal project, we developed a data-driven approach leveraging machine learning to identify patterns in supervisory data.
The patterns are then used as a basis for defining additional plausibility checks~\cite{subasps}.
Furthermore, it was of interest to identify individual observations in the reports which strongly deviated from the observed patterns.
In the context of this solution, we had to consider in particular the aspects of \emph{building trust}, \emph{gaining knowledge} and \emph{obtaining actionable insights}.

\subsubsection{Machine learning application}
Given the overall settings, the concrete objectives of our approach were to:

\begin{enumerate}[label=(\roman*)]
\item Identify novel functional dependencies between subsets of data points reported for supervisory data. 
\item Ensure the functional dependencies do not correspond to existing validation rules.
\item Present the functional dependency to business experts, such that they can analyze from a business perspective if the patterns motivate the introduction of new plausibility checks.
\item Identify anomalous values reported for supervisory data which motivate further checks on the level of individual data points.
\end{enumerate}

The solution we developed was based on running a multitude of regression analyses using Extremely Randomised Trees~\cite{geurts2006extremely}. 
The main steps to build a useful solution were to (i) ensure accurate results, (ii) provide interpretable representations of the patterns, and (iii) report outliers with a clear explanation of what rendered these data points suspicious \cite{subasps}.

The accuracy of the results involved several aspects. 
First, we ensured that only novel rules were detected.
To ensure that only novel rules are identified, information from existing validation rules was used to limit the search space for the pattern discovery, i.e. to exclude known functional dependencies when running the regression analysis.
This prevented the \emph{business experts} from being frustrated by a system that provided them with insights they already knew.
Second, we ranked the discovered patterns by their predictive quality. 
A transparent assessment of the Root Mean Squared Error of the prediction in a normalized value space helped to assess the quality of a pattern and ensured the business experts about the reliability of the functional dependence.
Finally, we filtered the ranked patterns by the number of influential variables in the regression analysis.
This helped to further focus on patterns that can be mapped more easily to business concepts.

The feature importance provided by the Extremely Randomised Trees implementation along with a transparent and methodologically clean quality assessment enabled us to address all these points. 
Using a non-linear transformation of the observed feature values to a standardized range rendered results comparable even if the original feature space differed by several orders of magnitude.
Using a transformation that was reversible further allowed to illustrate predicted values in the original value range and thereby again interpretable by the \emph{business experts}.

\subsubsection{Building Trust}
\emph{Business experts} are used to designing plausibility rules solely based on domain knowledge. 
In a user study, we presented the identified, ranked, and filtered patterns to \emph{business experts}, indicating the predictive quality in terms of Root Mean Squared Error and the list of influential variables.
The user study was run on a subset of the supervisory data the business experts were currently investigating for novel plausibility checks \cite{subasps}.
In the course of the study, the insights from identified patterns enabled the experts to define new plausibility checks. These checks have been implemented and are used to assess the quality of data.  This success story, based on xAI techniques created trust in the solution \cite{subasps}.

Furthermore, the \emph{data scientists} used the explanation for the identified patterns to test their solutions. By including the known validation rules in the search space, the approaches had to identify these rules as patterns in the data. This helped to create trust in the implementations provided by the \emph{data scientists}.

Finally, for \emph{business experts} and \emph{decision makers} it was of relevance to see that the approach is not biased. Given the heterogeneous landscape and business models of credit institutions, it was important that no particular group of institutions (e.g. big vs. small institutions) was particularly affected by the new checks.

\subsubsection{Gain Knowledge}
One of the main objectives of the solution was to identify new plausibility checks by gaining additional insights into the data \cite{subasps}. The detected patterns served as a kind of inspiration to look at new business motivated plausibility checks. 
Again the easily interpretable patterns with high predictive quality and few influential variables helped \emph{business experts} to connect the identified functional dependencies with their domain knowledge. The result was that the experts were inspired to detect new causal relationships in the data.

\subsubsection{Actionable Insights}
When analyzing and quality assuring supervisory data reported by a credit institution, the eventual aim is to detect anomalous values which require further investigations. 
An investigation, in this case, means that the \emph{business experts} contact the credit institution to ask for confirmation of the data and to ensure that no methodological or technical errors were made in the compilation and transmission of data. Hence, in this case the credit institutions correspond to the \emph{end user} being affected by an algorithmic decision.

With the investigated solution, this was achieved by means of comparing observed values vs. the forecasts made by the regression models.
Having converted the data to a normalized observation space provided comparable and standardized metrics of how far off an observation was from the expected value.
This provided the basis for a criticality assessment and a prioritization of outliers.
Furthermore, by mapping back the normalized values to the original data space, \emph{business experts} could not only interpret the results but also reach out to credit institutions and data compilers to ask for confirmation and clarification of the reported data, giving a clear indication of which value raised their attention and what they considered a plausible value range.

In a second user study, several suspicious values were presented to \emph{business experts}.
The \emph{business experts} investigated a few selected cases with credit institutions and in fact, they were able to communicate the cases.
This lead to the identification of cases of misreporting and the correction of data.

\section{Conclusions}\label{sec:conclusions}
In this paper, we described the desiderata, scope, and users of explainable machine learning needs in statistical production systems of the European Central Bank. 

We firstly introduced a summary of the different users and a classification of explainability needs that we have identified during our journey through the various machine learning projects. While the needs were distilled from our use cases in central banking, we believe they are of generic nature and the classification may serve for other use cases, as well.

Then we gave life to our classification by illustrating two different use cases that we analysed for explainability needs in the context of our experience with using machine learning for statistical production systems: data quality assurance at the Centralised Securities Database and identifying novel functional dependencies at the Supervisory Banking data system.

%We hope that our work facilitates the development of explainable AI methods that are incorporated into other real-world decision-making systems.

As future work, we intend to apply the classification of xAI desiderata to further use cases. We plan to use the classification framework for other use cases we envisage in the domain of central banking but are curious to expand also to other domains. Applying the classification scheme to more xAI use cases may lead to two developments: (i) a refinement of the classification scheme itself and (ii) enrichment of the entries in the scheme with references to suitable technical solutions.

\begin{acks}
This  work  was  partially  funded  by  the  European  Commission  under  contract numbers NoBIAS — H2020-MSCA-ITN-2019 project GA No. 860630. \\
\end{acks}

\bibliographystyle{ACM-Reference-Format}
\bibliography{sample}

\appendix

\end{document}